\documentclass[twocolumn,aps,superscriptaddress,showpacs,floatfix,longbibliography]{revtex4-1}
\usepackage{url}
\usepackage{cancel}
\usepackage[colorlinks,linkcolor=blue,citecolor=blue,filecolor=black,urlcolor=blue]{hyperref}
\usepackage{epsfig,graphics}
\usepackage{graphicx}
\usepackage{dcolumn}
\usepackage{bm}
\usepackage[usenames]{color}
\usepackage{ulem} 
\usepackage{amssymb}
\usepackage{amsmath}
\usepackage{multirow}
\usepackage{float}
\usepackage{MnSymbol}
\usepackage{appendix}
\usepackage{color}
\usepackage{cleveref}
\usepackage{mathtools}
\usepackage{dsfont} 

\begin{document}
\title{Bayesian inference of nucleus resonance and neutron skin}
\author{Jun Xu}\email{xujun@zjlab.org.cn}
\affiliation{School of Physics Science and Engineering, Tongji University, Shanghai 200092, China}
\affiliation{Shanghai Advanced Research Institute, Chinese Academy of Sciences, Shanghai 201210, China}
\affiliation{Shanghai Institute of Applied Physics, Chinese Academy of Sciences, Shanghai 201800, China}
\date{\today}

\begin{abstract}
In this proceeding, we have presented some highlight results on the constraints of the nuclear matter equation of state (EOS) from the data of nucleus resonance and neutron-skin thickness using the Bayesian approach based on the Skyrme-Hartree-Fock model and its extension. Typically, we have discussed the anti-correlation and positive correlation between the slope parameter and the value of the symmetry energy at the saturation density under the constraint of, respectively, the neutron-skin thickness and the isovector giant dipole resonance. We have shown that the Bayesian analysis can help to find a compromise for the ``PREXII puzzle'' and the ``soft Tin puzzle". We have further illustrated the possible modifications on the constraints of lower-order EOS parameters as well as the relevant correlation when higher-order EOS parameters are incorporated as independent variables. For a given model and parameter space, the Bayesian approach serves as a good analysis tool suitable for multi-messengers versus multi-variables, and is helpful for constraining quantitatively the model parameters as well as their correlations.
\end{abstract}
\maketitle

\section{Introduction}
\label{sec:intro}

Understanding properties of the nuclear interaction and the nuclear matter EOS is the basic goal of nuclear physics. Our knowledge on the nuclear matter EOS can be decoupled to the isospin-independent part and the isospin-dependent part, with the more uncertain latter part characterized by the nuclear symmetry energy $E_{sym}$. Extracting the density dependence of $E_{sym}$ has been a hot topic in the past twenty years, since it has important ramifications in nuclear structures, nuclear reactions, and nuclear astrophysics~\cite{Baran:2004ih,Steiner:2004fi,Lattimer:2006xb,Li:2008gp}. Properties of the single-nucleon potential is related to the macroscopic nuclear matter EOS and the microscopic nuclear interaction in the mean-field approximation, and it is momentum dependent in the non-relativistic case, as if nucleons propagate with an effective mass $m^\star$ in the nuclear medium (see Ref.~\cite{Li:2018lpy} for a recent review). Since both the nucleon effective mass and the nuclear matter EOS originate from the fundamental nuclear interaction, it is not surprising that they are related to each other through the Hugenholtz-Van Hove theorem~\cite{PhysRevC.82.054607,LI2013276}, while in many studies they are taken as independent quantities characterizing properties of the nuclear interaction in different aspects.

Nuclear resonances and neutron skins are accurate probes of the nuclear matter EOS and the nucleon effective mass around and below the saturation density. The isoscalar giant monopole resonance (ISGMR), a breathing mode of nucleons in the radial direction of the nucleus, is a famous probe of the incompressibility ($K_0$)~\cite{BLAIZOT1980171,PhysRevLett.82.691,GARG201855,Piekarewicz_2010,PhysRevLett.109.092501,Colo:2013yta,Shlomo,PhysRevC.97.025805} characterizing the isoscalar part of the EOS, while the isoscalar giant quadrupole resonance (ISGQR) is found to be more sensitive to the isoscalar nucleon effective mass~\cite{BLAIZOT1980171,doi:10.1142/3530,BOHIGAS1979267,PhysRevC.79.034310,PhysRevC.87.034301,PhysRevC.93.034335,PhysRevC.95.034324,PhysRevC.98.054316,PhysRevC.102.024306}. The isovector giant dipole resonance (IVGDR), an oscillation mode in which neutrons and protons move collectively relative to each other in a nucleus, is a good probe of the $E_{sym}$~\cite{Colo:2013yta,REINHARD1999305,PhysRevC.77.061304,PhysRevC.81.051303,PhysRevC.85.041302,PhysRevC.85.044317,PhysRevC.88.024316,PhysRevC.90.064317,PhysRevC.92.064304,PhysRevC.92.031301, PhysRevC.94.014313,PhysRevC.93.031301,PhysRevC.103.064301}, while it was recently found to be sensitive to the nucleon effective mass as well~\cite{PhysRevC.93.034335,PhysRevC.95.034324,PhysRevC.102.024306}. The neutron-skin thickness, defined as the difference in the root-mean-square radii of neutrons and protons, i.e., $\Delta r_{np} = \sqrt{\langle r_n^2 \rangle} - \sqrt{\langle r_p^2 \rangle}$, is one of the most robust probes of the slope parameter of the $E_{sym}$~\cite{PhysRevLett.85.5296,PhysRevC.64.027302,PhysRevLett.86.5647,FURNSTAHL200285,PhysRevLett.95.122501,PhysRevLett.102.122502,ZHANG2013234,Agrawal:2020wqj,Behera:2020wks,
PhysRevC.104.054324,PhysRevC.103.064323}. Despite the effectiveness of these probes in constraining the nuclear matter EOS and the nuclear interaction, contradictory information is sometimes obtained. For instance, the excitation energy of the ISGMR in Sn isotopes generally leads to a smaller incompressibility compared to that extracted from a heavy nucleus, leading to the famous ``soft Tin puzzle"~\cite{Piekarewicz_2010,PhysRevC.76.031301,GARG200736}. In addition, the recent PREXII experiments obtained a large $\Delta r_{np}$ in $^{208}$Pb through electron parity-violating scatterings~\cite{PhysRevLett.126.172502}, leading to a large slope parameter $L$ of the $E_{sym}$, contradictory to that extracted from the electric dipole polarizability of $^{208}$Pb~\cite{PhysRevC.104.024329} and the $\Delta r_{np}$ in $^{48}$Ca through the same experimental method~\cite{CREX:2022kgg}, and we call this the ``PREXII puzzle".

Since multipole observables (ISGMR, IVGDR, $\Delta r_{np}$) are sensitive to multipole physics quantities ($K_0$, $E_{sym}$, $m^\star$, etc.), it is proper to use the Bayesian approach in the study. The Bayesian analysis is helpful in not only constraining quantitatively the physics quantities but also obtaining their correlations under the constraints of experimental data. In our previous studies~\cite{XU2020135820,PhysRevC.102.044316,PhysRevC.104.054324,PhysRevC.105.044305}, we have extracted the information of the nuclear matter EOS and the nucleon effective mass from ISGMR, IVGDR, and $\Delta r_{np}$ through the Bayesian analysis based on the Skyrme-Hartree-Fock (SHF) model as well as its extension. In this proceeding, we present some of the highlight findings in our previous studies, including the negative and positive correlations between $L$ and $E_{sym}$ at the saturation density from $\Delta r_{np}$ and IVGDR, respectively, the way to get compatible results from ``conflict" data, and the correlation between lower-order and higher-order EOS parameters.

\section{Theoretical framework}
\label{sec:theory}

In this section, we briefly review the theoretical framework for the series of studies on the Bayesian inference of nucleus resonance and neutron skin. We will review briefly the definition of the EOS parameters and the nucleon effective mass, the SHF model and its extension, the random-phase approximation method for nucleus resonance, and the main formulaes of the Bayesian approach.

\subsection{Definition of EOS parameters and effective mass}

The binding energy per nucleon in isospin asymmetric nuclear matter with nucleon density $\rho=\rho_n+\rho_p$ and isospin asymmetry $\delta = (\rho_n-\rho_p)/\rho$ can be expressed as
\begin{equation}
E(\rho,\delta) = E_0(\rho) + E_{sym}(\rho) \delta^2 + O(\delta^4).
\end{equation}
where the symmetry energy is defined as
\begin{equation}
E_{sym}(\rho) = \frac{1}{2} \left[\frac{\partial^2 E(\rho,\delta)}{\partial \delta^2}\right]_{\delta=0}.
\end{equation}
The higher-order $\delta$ terms are generally much smaller than the lower-order terms, so the EOS is mostly dominated by $E_0(\rho)$ and $E_{sym}(\rho)$.

Around the saturation density $\rho_0$, $E_0(\rho)$ and $E_{sym}(\rho)$ can be expanded in the power of $\chi = (\rho-\rho_0)/{3\rho_0}$ as
\begin{eqnarray}
E_0(\rho) &=& E_0(\rho_0) + \frac{K_0}{2!} \chi^2 + \frac{Q_0}{3!} \chi^3 + O(\chi^4), \nonumber \\
E_{sym}(\rho) &=& E_{sym}(\rho_0) + L \chi + \frac{K_{sym}}{2!} \chi^2 + \frac{Q_{sym}}{3!} \chi^3 + O(\chi^4). \nonumber
\end{eqnarray}
In the above, the linear term in the expansion of $E_0(\rho)$ vanishes due to zero pressure of symmetric nuclear matter (SNM) at $\rho_0$. The EOS parameters relevant in the study are physics quantities at $\rho_0$ including the isoscalar ones, i.e., the incompressibility $K_0$ and the skewness parameter $Q_0$ of the SNM EOS, and the isovector ones, i.e., the value $E_{sym}^0$ and the slope parameter $L$, the curvature parameter $K_{sym}$, and the skewness parameter $Q_{sym}$ of the symmetry energy, and they are defined respectively as
\begin{eqnarray}
&&K_0 \equiv 9\rho_0^2 \left[\frac{\partial^2 E_{SNM}(\rho)}{\partial \rho^2}\right]_{\rho=\rho_0},\\
&&Q_0 \equiv 27\rho_0^3 \left[\frac{\partial^3 E_{SNM}(\rho)}{\partial \rho^3}\right]_{\rho=\rho_0},\\
&&E_{sym}^0 \equiv E_{sym}(\rho_0),\\
&&L \equiv 3\rho_0 \left[\frac{\partial E_{sym}(\rho)}{\partial \rho}\right]_{\rho=\rho_0},\\
&&K_{sym} \equiv 9\rho_0^2 \left[\frac{\partial^2 E_{sym}(\rho)}{\partial \rho^2}\right]_{\rho=\rho_0},\\
&&Q_{sym} \equiv 27\rho_0^3 \left[\frac{\partial^3 E_{sym}(\rho)}{\partial \rho^3}\right]_{\rho=\rho_0}.
\end{eqnarray}

The p-mass of nucleons with isospin index $\tau$ in the non-relativistic model is related to the momentum dependence of the single-nucleon potential $U_\tau$, i.e.,
\begin{equation}
\frac{m^\star_\tau}{m} = \left( 1+ \frac{m}{p}\frac{dU_\tau}{dp}\right)^{-1},
\end{equation}
with $m$ being the bare nucleon mass. The effective mass of neutrons or protons depends on the nucleon momentum as well as the density and isospin asymmetry of the nuclear medium, but generally represented by the value at the Fermi momentum in normal nuclear matter. The isoscalar effective mass $m_s^\star$ is the nucleon effective mass in SNM, and the isovector effective mass $m_v^\star$ is the neutron (proton) effective mass in pure proton (neutron) matter.

\subsection{Skyrme-Hartree-Fock model and its extension}

The effective interaction between nucleons at $\vec{r}_1$ and $\vec{r}_2$ in the standard SHF model can be expressed as
\begin{eqnarray}\label{v12}
v^{SHF}(\vec{r}_1,\vec{r}_2) &=& t_0(1+x_0P_\sigma)\delta(\vec{r}) \notag \\
&+& \frac{1}{2} t_1(1+x_1P_\sigma)[{\vec{k}'^2}\delta(\vec{r})+\delta(\vec{r})\vec{k}^2] \notag\\
&+&t_2(1+x_2P_\sigma)\vec{k}' \cdot \delta(\vec{r})\vec{k} \notag\\
&+&\frac{1}{6}t_3(1+x_3P_\sigma)\rho^\alpha(\vec{R})\delta(\vec{r})\notag\\
&+& i W_0(\vec{\sigma}_1+\vec{\sigma_2})[\vec{k}' \times \delta(\vec{r})\vec{k}].
\end{eqnarray}
In the above, $\vec{r}=\vec{r}_1-\vec{r}_2$ and $\vec{R}=(\vec{r}_1+\vec{r}_2)/2$ are respectively the relative and central coordinates for the two nucleons, $\vec{k}=(\nabla_1-\nabla_2)/2i$ is the relative momentum operator and $\vec{k}'$ is its complex conjugate acting on the left, and $P_\sigma=(1+\vec{\sigma}_1 \cdot \vec{\sigma}_2)/2$ is the spin exchange operator. The spin-orbit coupling constant is fixed at its default value $W_0=133$ MeVfm$^5$. The parameters $t_0$, $t_1$, $t_2$, $t_3$, $x_0$, $x_1$, $x_2$, $x_3$, and $\alpha$ can be solved inversely from the macroscopic quantities~\cite{Chen:2010qx}, i.e., the saturation density $\rho_0$, the binding energy $E_0$ at the saturation density, the incompressibility $K_0$, the isoscalar and isovector nucleon effective mass $m_s^\star$ and $m_v^\star$, the symmetry energy $E_{sym}^0$ and its slope parameter $L$ at the saturation density, and the isoscalar and isovector density gradient coefficient $G_S$ and $G_V$.

The above SHF model has many extensions, while the Korea-IBS-Daegu-SKKU (KIDS) model~\cite{PhysRevC.97.014312,PhysRevC.99.064319} is one of them, by replacing the density-dependent term in the effective interaction [Eq.~(\ref{v12})] with the following form
\begin{equation}
v^{KIDS}_\rho (\vec{r}_1,\vec{r}_2) = \frac{1}{6}\sum_{i=1}^{3} (t_{3i}+y_{3i}P_\sigma)\rho^{i/3}(\vec{R})\delta(\vec{r}).
\end{equation}
Compared to the standard SHF model, the additional coefficients in the KIDS model, i.e., $t_{3i}$ and $y_{3i}$, allow us to vary more individual EOS parameters, i.e., $Q_0$, $K_{sym}$, and $Q_{sym}$ as shown in Ref.~\cite{PhysRevC.105.044305}.

The energy-density functional form in the SHF model as well as its extension can be derived based on the Hartree-Fock method, and the single-nucleon Hamiltonian can then be obtained with the variational principle. Solving the Shr\"odinger equation leads to the wave function of each nucleon as well as the density distributions for spherical nuclei, and thus the neutron-skin thickness. For the Hartree-Fock calculation, we use the open source code described in Ref.~\cite{Reinhard1991}.

\subsection{Random-phase approximation method for nucleus resonance}

The nuclear response to external fields is studied by applying the random phase approximation (RPA) and using the Hartree-Fock basis. In the studies relevant to nucleus resoance, we use the open source routine of Ref.~\cite{COLO2013142} with certain modifications. The operators for exciting the IVGDR, ISGMR, and ISGQR are chosen respectively as
\begin{equation}
\hat{F}_{\rm IVGDR} = \frac{N}{A} \sum_{i=1}^Z r_i Y_{\rm 1M}(\hat{r}_i) - \frac{Z}{A} \sum_{i=1}^N r_i Y_{\rm 1M}(\hat{r}_i), \label{QIVGDR}
\end{equation}
\begin{equation}
\hat{F}_{\rm ISGMR} = \sum_{i=1}^A r_i^2 Y_{00}(\hat{r}_i),
\end{equation}
\begin{equation}
\hat{F}_{\rm ISGQR} = \sum_{i=1}^A r_i^2 Y_{2M}(\hat{r}_i),
\end{equation}
where $N$, $Z$, and $A$ are respectively the neutron, proton, and nucleon numbers in a nucleus, $r_i$ is the coordinate of the $i$th nucleon with respect to the center-of-mass of the nucleus, and $Y_{\rm 00}(\hat{r}_i)$, $Y_{\rm 1M}(\hat{r}_i)$, and $Y_{2M}(\hat{r}_i)$ are the spherical harmonics with the magnetic quantum number $M$ degenerate in spherical nuclei. Using the RPA method~\cite{COLO2013142}, the strength function
\begin{equation}
S(E) = \sum_\nu |\langle \nu|| \hat{F}  || \tilde{0} \rangle |^2 \delta(E-E_\nu)
\end{equation}
of a nucleus resonance in a given channel can be obtained, where the square of the reduced matrix element $|\langle \nu|| \hat{F}  || \tilde{0} \rangle |$ represents the transition probability from the ground state $| \tilde{0} \rangle $ to the excited state $| \nu \rangle$ under the action of the external field $\hat{F}$. The moments of the strength function for the corresponding resonance type can then be calculated from
\begin{equation}
m_k = \int_0^\infty dE E^k S(E).
\end{equation}
For the IVGDR, the centroid energy $E_{-1}$ and the electric polarizability $\alpha_D$ can be obtained from the moments of the strength function through the relation
\begin{eqnarray}
E_{-1} &=& \sqrt{m_1/m_{-1}}, \\
\alpha_D &=& \frac{8\pi e^2}{9} m_{-1}.
\end{eqnarray}
For the ISGMR, the RPA results of the excitation energy
\begin{equation}
E_{ISGMR}=m_1/m_0
\end{equation}
are compared with the corresponding experimental data. For the ISGQR, we compare the peak values of the strength function directly to the corresponding experimental data, and the value of $m_s^\star$ is determined in this way.

\subsection{Bayesian analysis}

We employ the Bayesian analysis method to obtain the probability distribution functions (PDFs) of model parameters from the experimental data, and the calculation method can be formally expressed as the Bayes' theorem
\begin{equation}
P(M|D) = \frac{P(D|M)P(M)}{\int P(D|M)P(M)dM},
\end{equation}
where $P(M|D)$ is the posterior probability for the model $M$ given the data set $D$, $P(D|M)$ is the likelihood function or the conditional probability for a given theoretical model $M$ to predict correctly the data $D$, and $P(M)$ denotes the prior probability of the model $M$ before being confronted with the data. The denominator of the right-hand side of the above equation is the normalization constant. For the prior PDFs, different combinations of model parameters $p_1=E_{sym}^0$, $p_2=L$, $p_3=m_v^\star/m$, $p_4=K_0$, $p_5=K_{sym}$, $p_6=Q_{sym}$, and $p_7=Q_0$ are varied uniformly within their empirical ranges. The theoretical results of $d^{th}_1=\Delta r_{np}$, $d^{th}_2=E_{-1}$, $d^{th}_3=\alpha_D$, and $d^{th}_4=E_{ISGMR}$ from the SHF-RPA method are compared with the experimental data $d^{exp}_{1 \sim 4}$, and a likelihood function is used to quantify how well these model parameters reproduce the corresponding experimental data
\begin{eqnarray}
&&P[D(d_1,d_2,d_3,d_4)|M(p_1,p_2,p_3,p_4,p_5,p_6,p_7)] \notag\\
&=& \Pi_{i=1}^4 \Bigg \{ \frac{1}{2\pi \sigma_i} \exp\left[-\frac{(d^{th}_i-d^{exp}_i)^2}{2\sigma_i^2}\right] \Bigg\}, \label{llh}
\end{eqnarray}
where $\sigma_{i}$ is the $1\sigma$ error of the data $d_i^{exp}$. The calculation of the posterior PDFs is based on the Markov-Chain Monte Carlo (MCMC) approach using the Metropolis-Hastings algorithm. Since the MCMC process does not start from an equilibrium distribution, initial samples in the so-called burn-in period have to be thrown away. After the average of each model parameter becomes stable, the posterior PDF of a single model parameter $p_i$ can be calculated from
\begin{equation}\label{1dpdf}
P(p_i|D) = \frac{\int P(D|M) P(M) \Pi_{j\ne i} dp_j}{\int P(D|M) P(M) \Pi_{j} dp_j},
\end{equation}
while the correlated PDF of two model parameters $p_i$ and $p_j$ can be calculated from
\begin{equation}\label{2dpdf}
P[(p_i,p_j)|D] = \frac{\int P(D|M) P(M) \Pi_{k\ne i,j} dp_{k}}{\int P(D|M) P(M) \Pi_{k} dp_k}.
\end{equation}
For the one-dimensional PDF, the range of the model parameter at the $68\%$ confidence level is obtained according to
\begin{equation}
\int_{p_{iL}}^{p_{iU}}  P(p_i|D) dp_i = 0.68, \label{confidence}
\end{equation}
where $p_{iL}$ ($p_{iU}$) is the lower (upper) limit of the corresponding narrowest interval of the parameter $p_i$ surrounding its mean value or its maximum {\it a posteriori} (MAP) value.

\section{Highlight results and discussions}
\label{sec:results}

With the theoretical framework given above, we have obtained the PDFs and correlations for various EOS parameters under the constraints of nucleus resonances and neutron-skin thicknesses mostly in $^{208}$Pb and Sn isotopes based on the Bayesian approach~\cite{XU2020135820,PhysRevC.102.044316,PhysRevC.104.054324,PhysRevC.105.044305}. In this section, we present some highlight results from previous studies.

\subsection{Correlations between $L$ and $E_{sym}^0$}

\begin{figure}[ht]
	\includegraphics[scale=0.35]{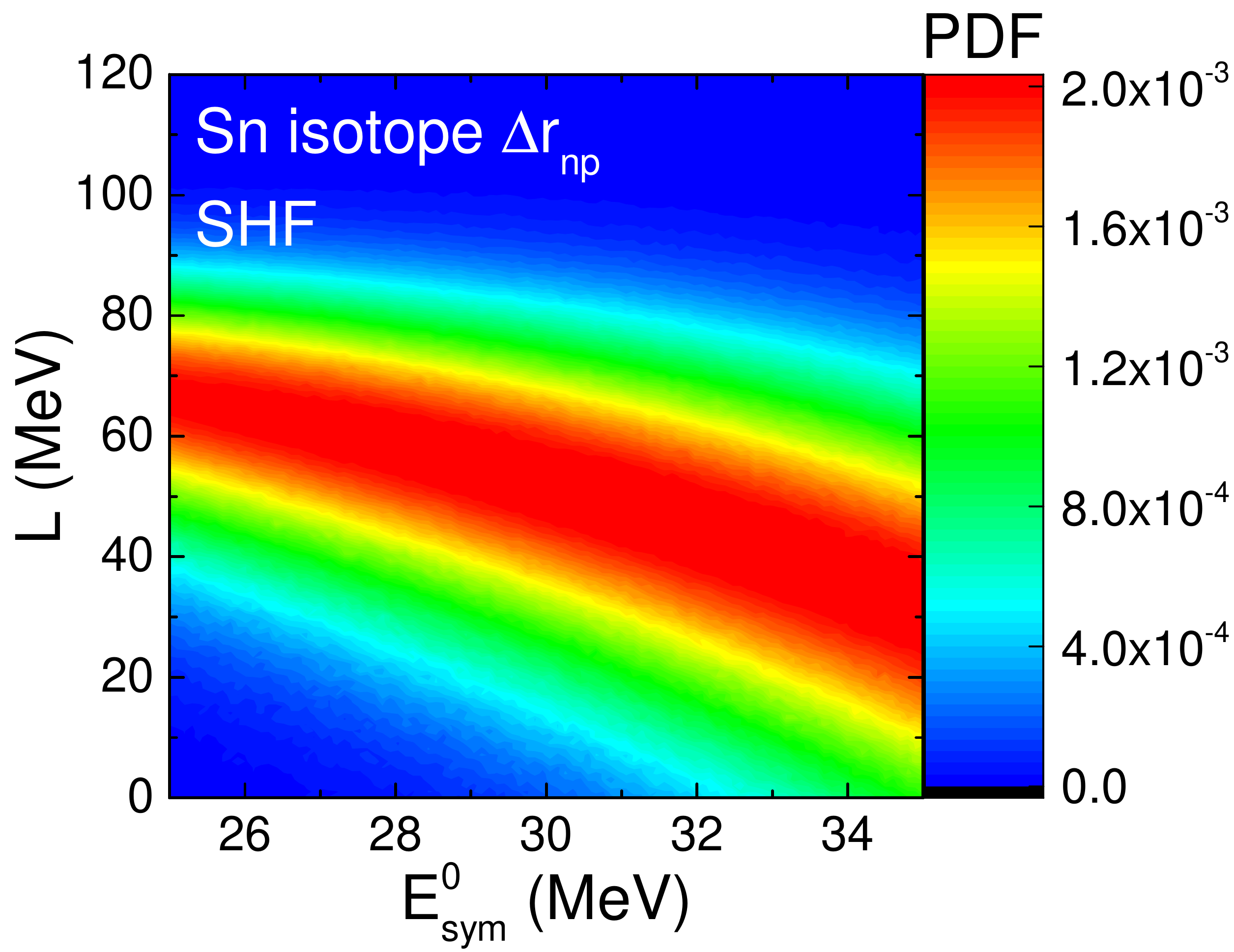}
	\caption{(Color online) Correlated PDF between $L$ and $E_{sym}^0$ from the Bayesian inference of the neutron-skin thickness in Sn isotopes based on the standard SHF model. Taken from Ref.~\cite{PhysRevC.102.044316}.}\label{L0-Esym0}
\end{figure}

\begin{figure}[ht]
	\includegraphics[scale=0.3]{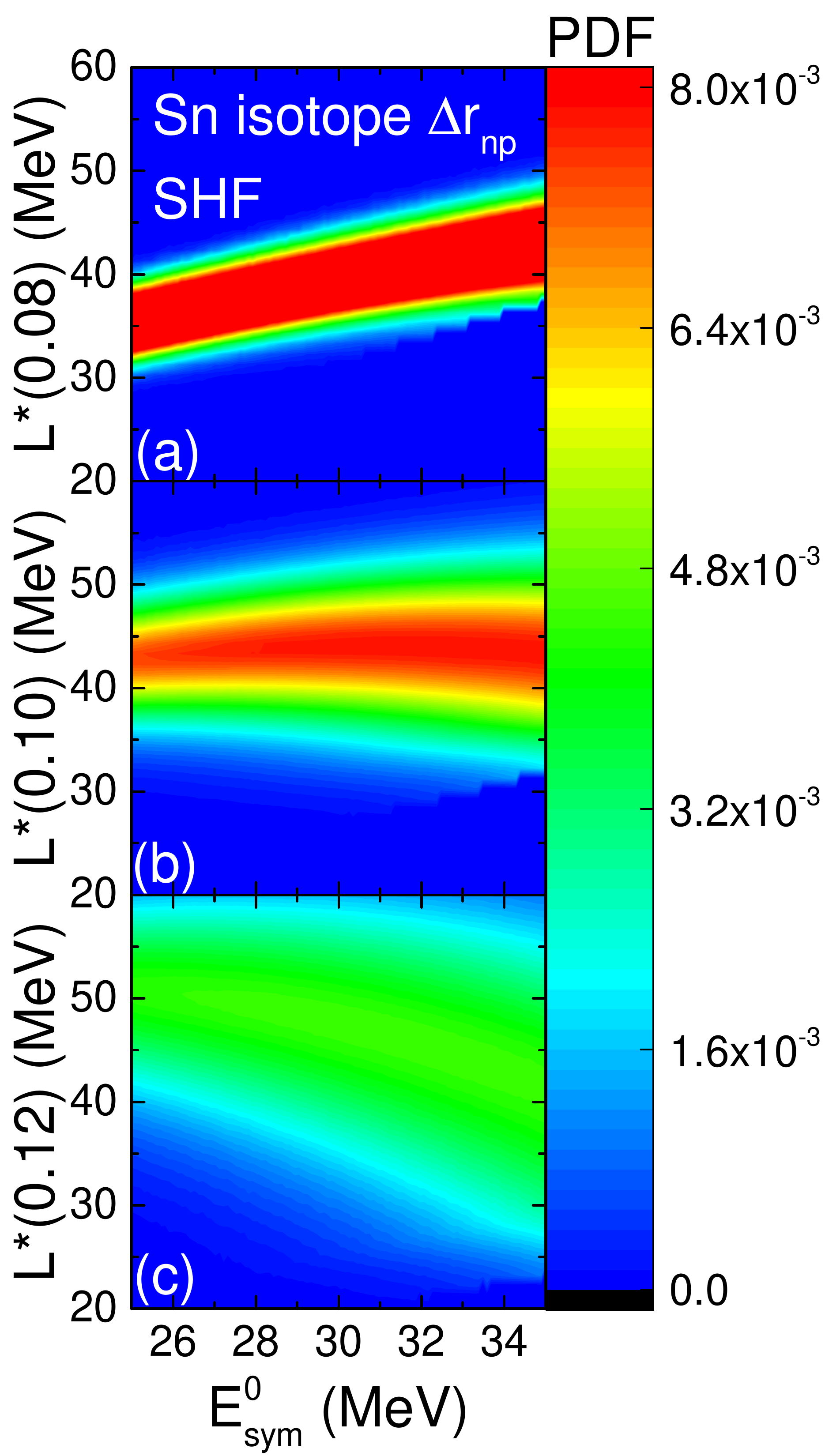}
	\caption{(Color online) Correlated PDFs between the slope parameter at 0.08 (a), 0.10 (b), and 0.12 (c) fm$^{-3}$ and $E_{sym}^0$ from the Bayesian inference of the neutron-skin thickness in Sn isotopes based on the standard SHF model. Taken from Ref.~\cite{PhysRevC.102.044316}.}\label{L-Esym0}
\end{figure}

\begin{figure}[ht]
	\includegraphics[scale=0.3]{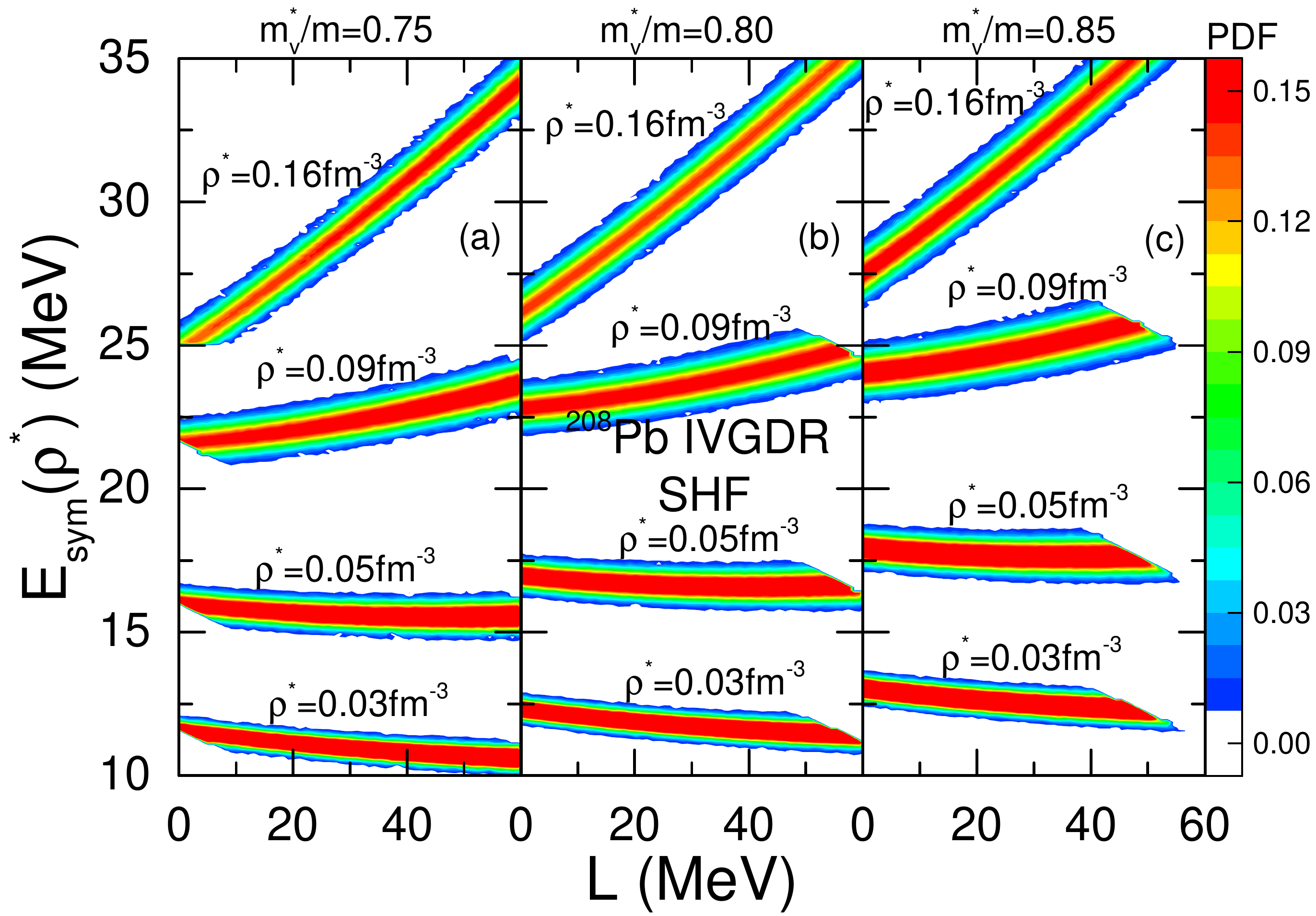}
	\caption{(Color online) Correlated PDFs between the symmetry energy at different densities and $L$ for $m_v^\star/m=0.75$ (a), 0.80 (b), and 0.85 (c) from the Bayesian inference of the IVGDR data for $^{208}$Pb based on the standard SHF model. Taken from Ref.~\cite{XU2020135820}.}\label{Esym0-L0}
\end{figure}

The correlation between $L$ and $E_{sym}^0$ has attracted considerable interest, and its information can be helpful for understanding the density dependence of the symmetry energy. As shown in Fig.~1 of Ref.~\cite{Lattimer:2014sga}, most isovector probes lead to positive correlations between $L$ and $E_{sym}^0$, while a negative correlation between $L$ and $E_{sym}^0$ was obtained under the constraint of the neutron-skin thickness $\Delta r_{np}$ in Sn isotopes from the $\chi^2$ fit by Ref.~\cite{Chen:2010qx}, so this needs some further discussions. As shown in Fig.~\ref{L0-Esym0}, we have obtained a similar anti-correlation between $L$ and $E_{sym}^0$ as in Ref.~\cite{Chen:2010qx} under the constraint of the $\Delta r_{np}$ in Sn isotopes but based on the Bayesian analysis. To further explore the origin of this anti-correlation, we have calculated the correlation between the slope parameter $L^\star = 3\rho^\star \left(dE_{sym}/d\rho\right)_{\rho^\star}$ at different subsaturation densities $\rho^\star$ with $E_{sym}^0$, and the results are displayed in Fig.~\ref{L-Esym0}. For $\rho^\star<0.10$ fm$^{-3}$, one observes a positive correlation between $L^\star$ and $E_{sym}^0$. For $\rho^\star>0.10$ fm$^{-3}$, a negative correlation between $L^\star$ and $E_{sym}^0$ is seen. At $\rho^\star=0.10$ fm$^{-3}$, which is approximately the average density of a nucleus, $L^\star$ is roughly independent of $E_{sym}^0$. As first pointed out in Ref.~\cite{ZHANG2013234}, this shows that the $\Delta r_{np}$ mostly constrains the value of $L^\star$ at $\rho^\star=0.10$ fm$^{-3}$, while its constraining power on the $E_{sym}(\rho)$ away from $\rho=\rho^\star$ is reduced and depends on the energy-density functional. As illustrated in the appendix of Ref.~\cite{PhysRevC.102.044316}, we have shown with a simply parameterized symmetry energy $E_{sym}(\rho)=E_{sym}^0(\rho/\rho_0)^\gamma$ that a negative correlation $L$ and $E_{sym}^0$ can be obtained if the value of $L^\star$ at a certain subsaturation density $\rho^\star$ is constrained.

The $\Delta r_{np}$ is roughly independent of the nucleon effective mass, while the spectrum of the nucleus resonance may depend on the nucleon effective mass. With the isoscalar nucleon effective mass fixed by the ISGQR data, the IVGDR data may help to constrain the PDFs of $L$, $E_{sym}^0$, and $m_v^\star$. For different values of $m_v^\star/m$, Fig.~\ref{Esym0-L0} displays the correlated PDFs between $E_{sym}(\rho^\star)$ and $L$ at different densities $\rho^\star$ under the constraint of the IVGDR in $^{208}$Pb. At $\rho^\star=\rho^0$, a positive correlation between $E_{sym}^0$ and $L$ is observed. One can further see that the correlation between $E_{sym}(\rho^\star)$ and $L$ is positive for $\rho^\star>0.05$ fm$^{-3}$ but negative for $\rho^\star<0.05$ fm$^{-3}$. At around $\rho^\star=0.05$ fm$^{-3}$, $E_{sym}(\rho^\star)$ becomes approximately independent of $L$. As first pointed out in Ref.~\cite{PhysRevC.90.064317}, this shows that the IVGDR data mostly constrains the value of $E_{sym}(\rho^\star)$ at $\rho^\star=0.05$ fm$^{-3}$. As also illustrated in the appendix of Ref.~\cite{PhysRevC.102.044316}, we have shown that a positive correlation between $E_{sym}^0$ and $L$ can be obtained if the value of $E_{sym}(\rho^\star)$ at a certain subsaturation density $\rho^\star$ is constrained.

\subsection{Compatibility between ``conflict'' data}

\begin{figure}[ht]
	\includegraphics[scale=0.3]{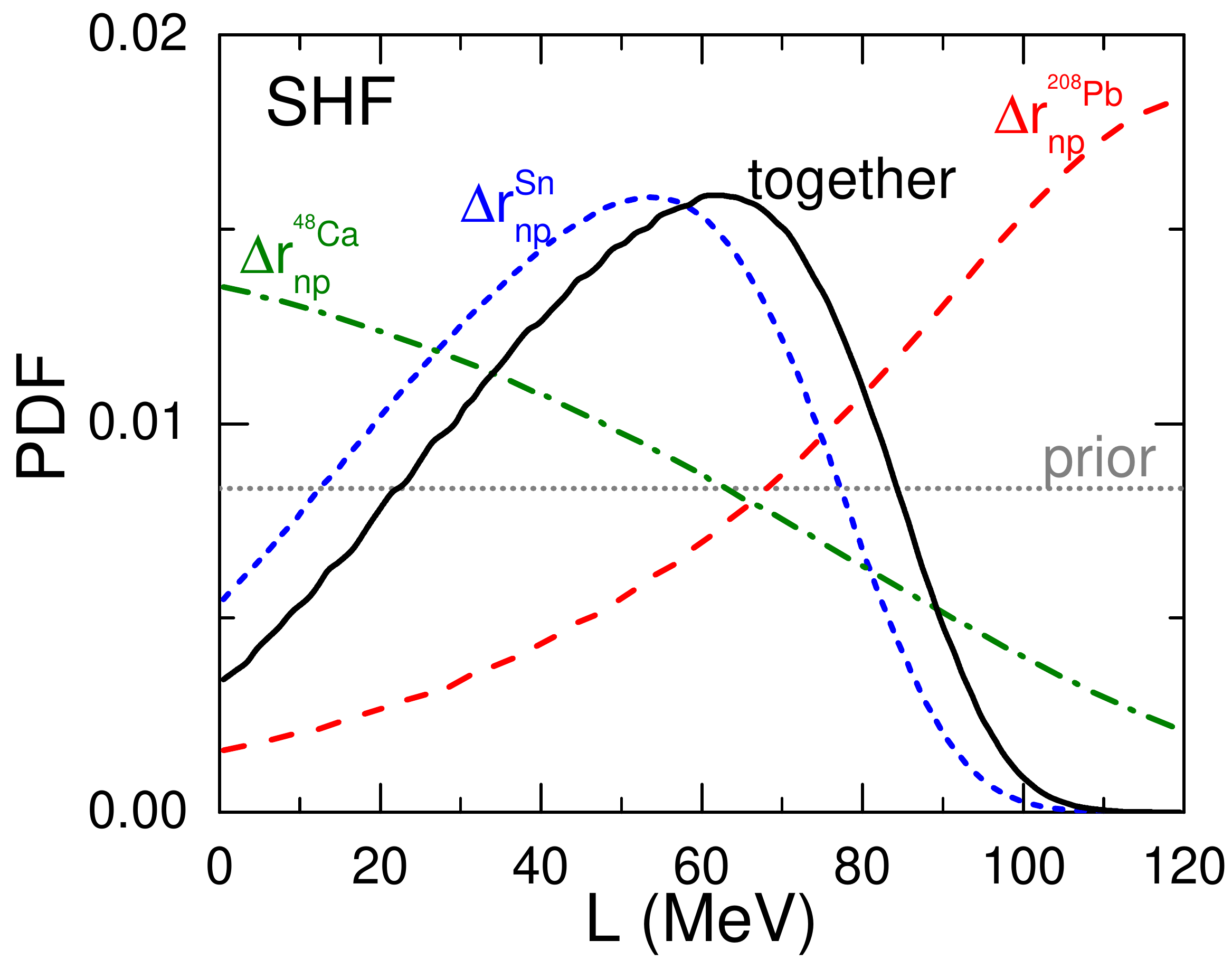}
	\caption{(Color online) Posterior PDFs of $L$ from the Bayesian inference of the neutron-skin thickness in Sn isotopes, $^{208}$Pb, $^{48}$Ca, and them together based on the standard SHF model.}\label{L}
\end{figure}

\begin{figure}[ht]
	\includegraphics[scale=0.3]{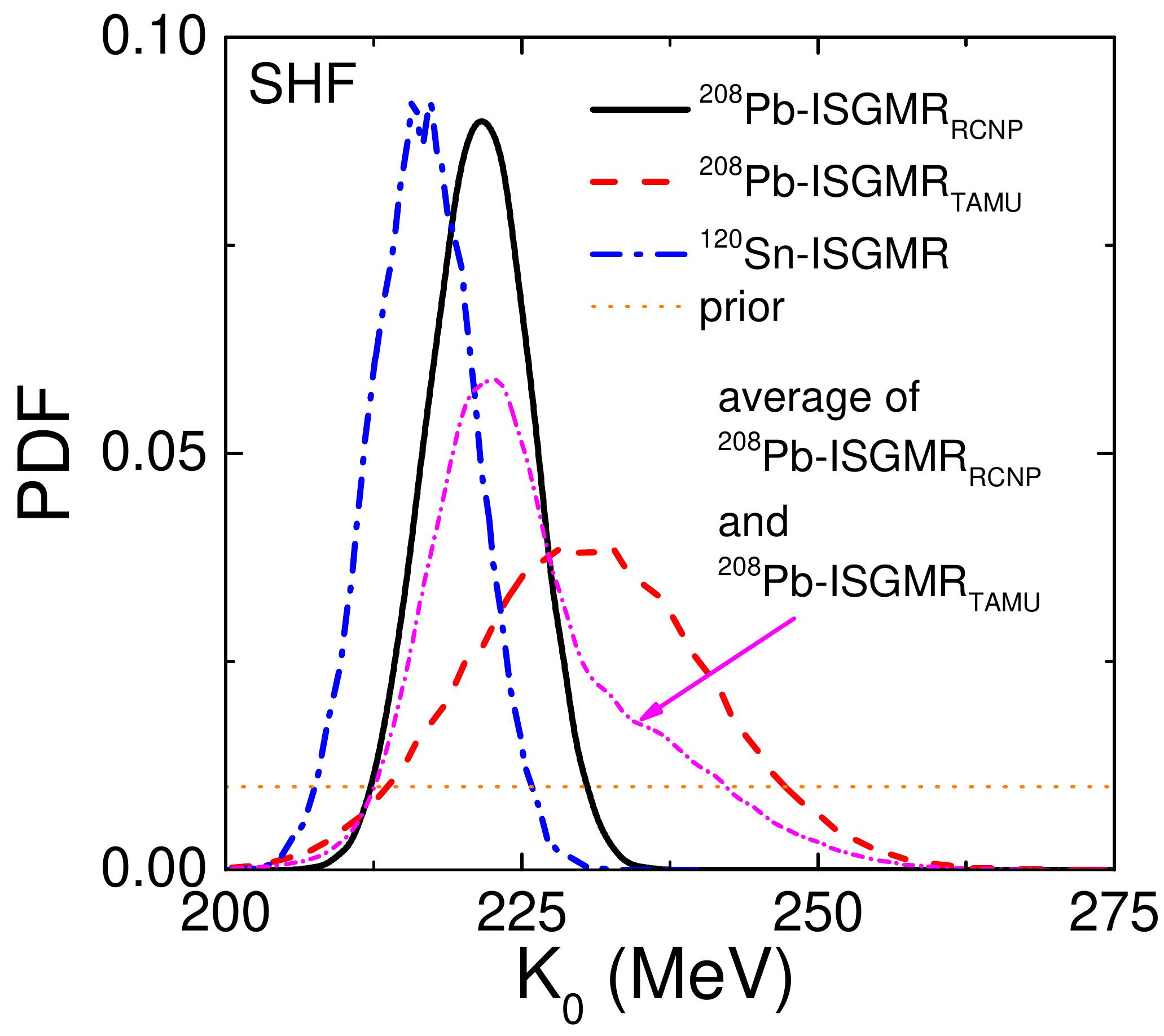}
	\caption{(Color online) Posterior PDFs of $K_0$ from the Bayesian inference of the nuclear structure data of $^{208}$Pb including the ISGMR data from RCNP and TAMP as well as that of $^{120}$Sn based on the standard SHF model. Taken from Ref.~\cite{PhysRevC.104.054324}.}\label{K0}
\end{figure}

The Bayesian analysis can also help to reach a compatibility between ``conflict'' data. From parity-violating electron scatterings, the measured neutron-skin thickness is $\Delta r_{np}=0.283 \pm 0.071$ fm for $^{208}$Pb~\cite{PhysRevLett.126.172502} and $\Delta r_{np}=0.121 \pm 0.026 (model) \pm 0.024 (theo)$ fm for $^{48}$Ca~\cite{CREX:2022kgg}. While the central values of the experimental data favor a large and small $L$, respectively, there are still chances to make them compatible since the error bars are also large. Taking the $1\sigma$ error from the experimental measurement as the width in the likelihood function [Eq.~(\ref{llh})], the PDFs of $L$ from the Bayesian inference of the $\Delta r_{np}$ in $^{208}$Pb and $^{48}$Ca are compared to that from the $\Delta r_{np}$ in Sn isotopes in Fig.~\ref{L}. Due to the large error bars for the $\Delta r_{np}$ in $^{208}$Pb and $^{48}$Ca, the corresponding PDFs are very broad, though they peak at a rather large and small value, respectively, compared to the PDF of $L$ from the $\Delta r_{np}$ in Sn isotopes. This shows that the constraining power of the experimental data is weaker with a large error bar. The considerable overlap of the PDFs of $L$ shows the compatibility of the $\Delta r_{np}$ data set for $^{208}$Pb, $^{48}$Ca, and Sn isotopes. We have further obtained the PDF of $L$ by including all the $\Delta r_{np}$ data, and this leads to a PDF of $L$ similar to that from the $\Delta r_{np}$ in Sn isotopes. Within $68\%$ confidence level, this PDF gives $L=51.6^{+28.7}_{-19.9}$ MeV surrounding its mean value or $L=61.5^{+18.8}_{-29.8}$ MeV surrounding its MAP value. As for the conflict between the $\Delta r_{np}$ data and the IVGDR data for $^{208}$Pb, since the IVGDR data constrains $E_{sym}(\rho^\star)$ at $\rho^\star=0.05$ fm$^{-3}$ and the nucleon effective mass, as shown in Ref.~\cite{XU2020135820}, there is no directly conflict on the PDF of $L$ from our point of view. Based on the above discussions, a compromise for the ``PREXII puzzle" can be found.

We now turn to the famous ``soft Tin puzzle''. With the isovector EOS parameter constrained by the $\Delta r_{np}$ and IVGDR data, the resulting PDFs of $K_0$ from the ISGMR data for $^{208}$Pb and $^{120}$Sn are compared in Fig.~\ref{K0}. The difference in the excitation energy of the ISGMR in $^{208}$Pb from TAMU and RCNP leads to different MAP values of the $K_0$, while they are both larger than the MAP value of the $K_0$ from the ISGMR data of $^{120}$Sn. This is the puzzle ``why Tin is so soft''. Without modifying the basic theoretical SHF-RPA framework, one observes considerable overlap in the PDFs of $K_0$ from both cases, showing that we can find a compromise for the ``soft Tin puzzle'' as well.

\subsection{Correlation between lower-order and higher-order EOS parameters}

The calculations in the previous subsections are all based on the standard SHF model, where only lower-order EOS parameters are varied, while higher-order EOS parameters are simultaneously changed according to the energy-density functional. In the KIDS model, we are able to vary independently both lower-order and higher-order EOS parameters, with the latter including $Q_0$, $K_{sym}$, and $Q_{sym}$. Since higher-order EOS parameters generally have larger empirical uncertainty ranges, the constraints on lower-order EOS parameters may change in this case.

\begin{figure}[ht]
	\includegraphics[scale=0.16]{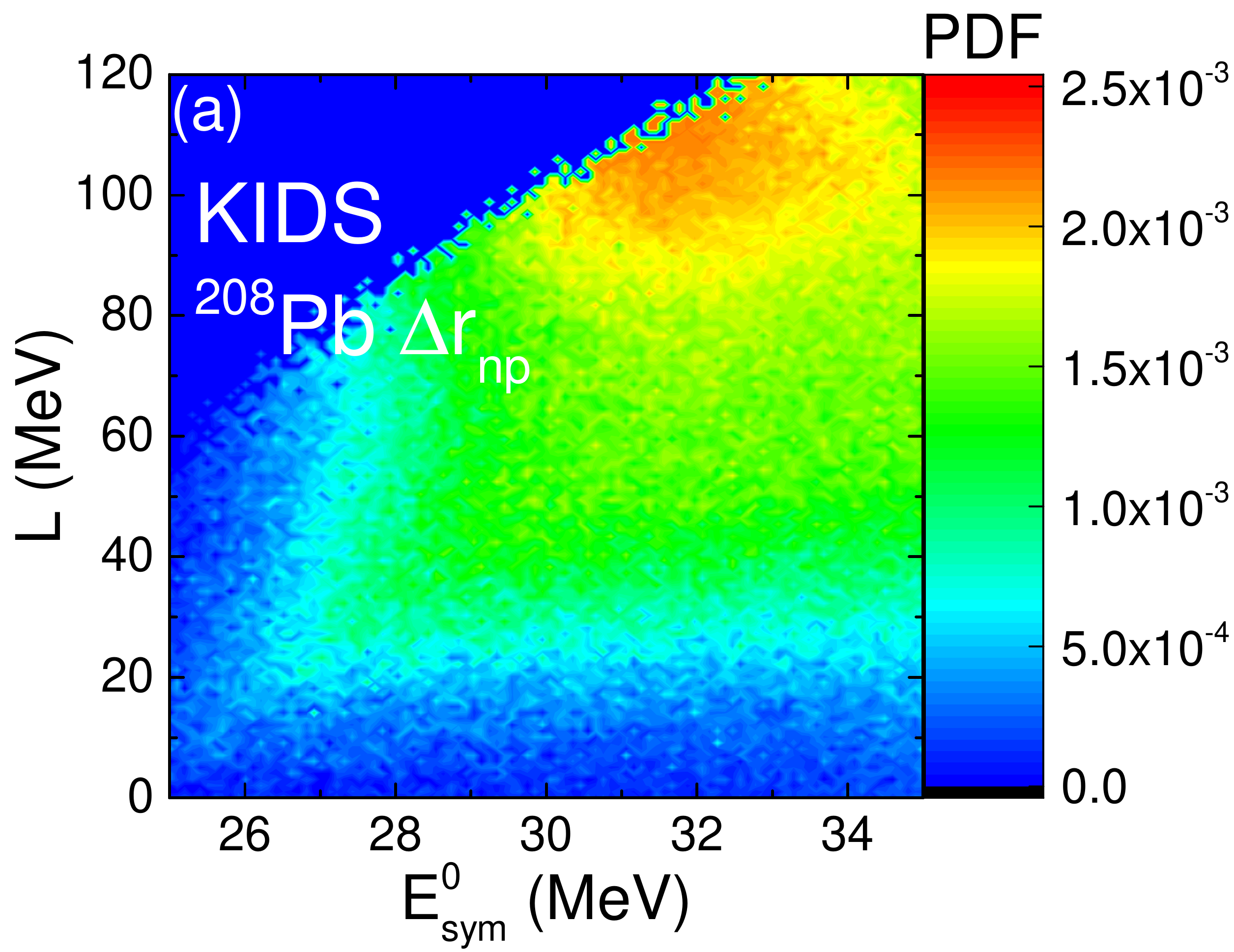}
	\includegraphics[scale=0.16]{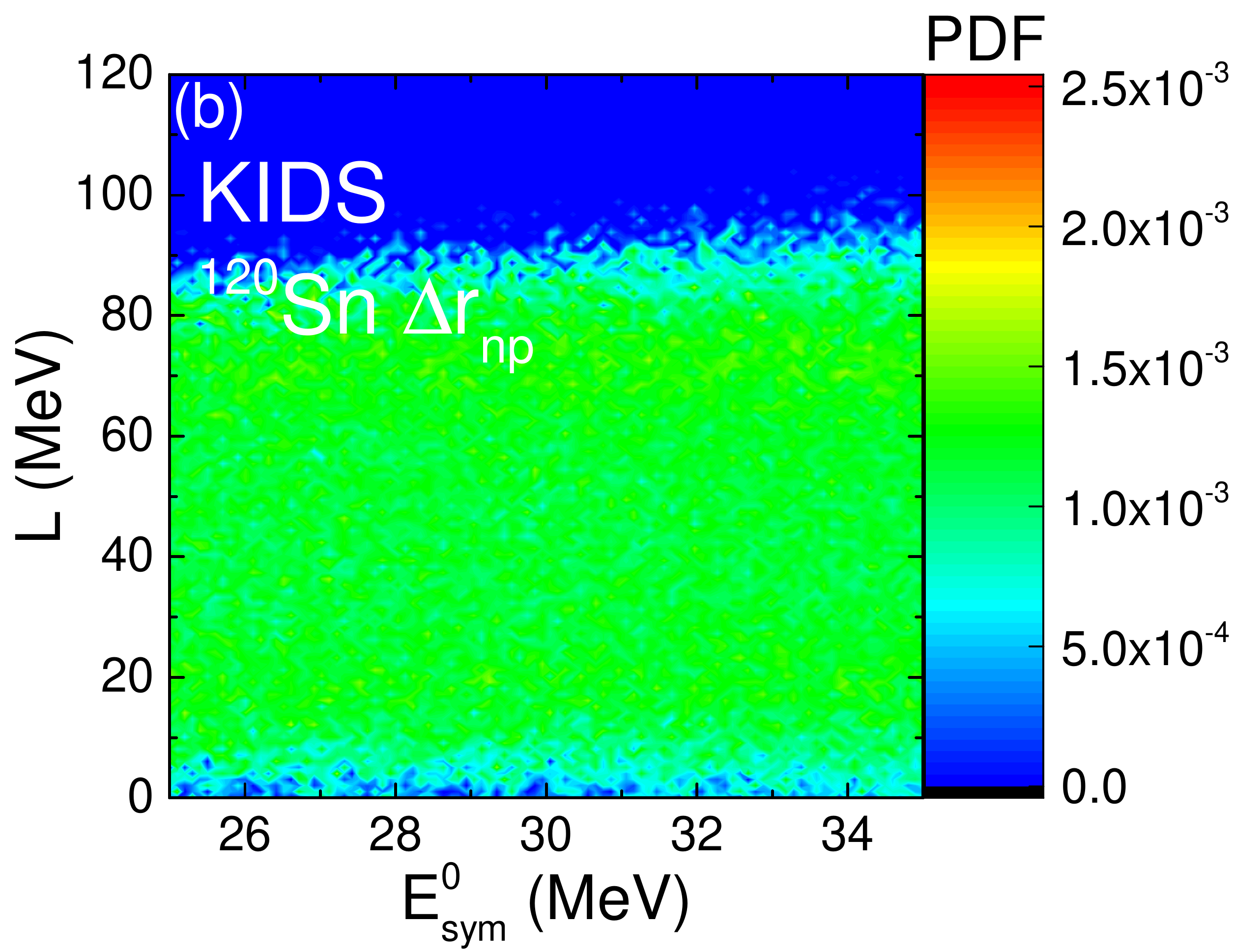}\\
	\includegraphics[scale=0.16]{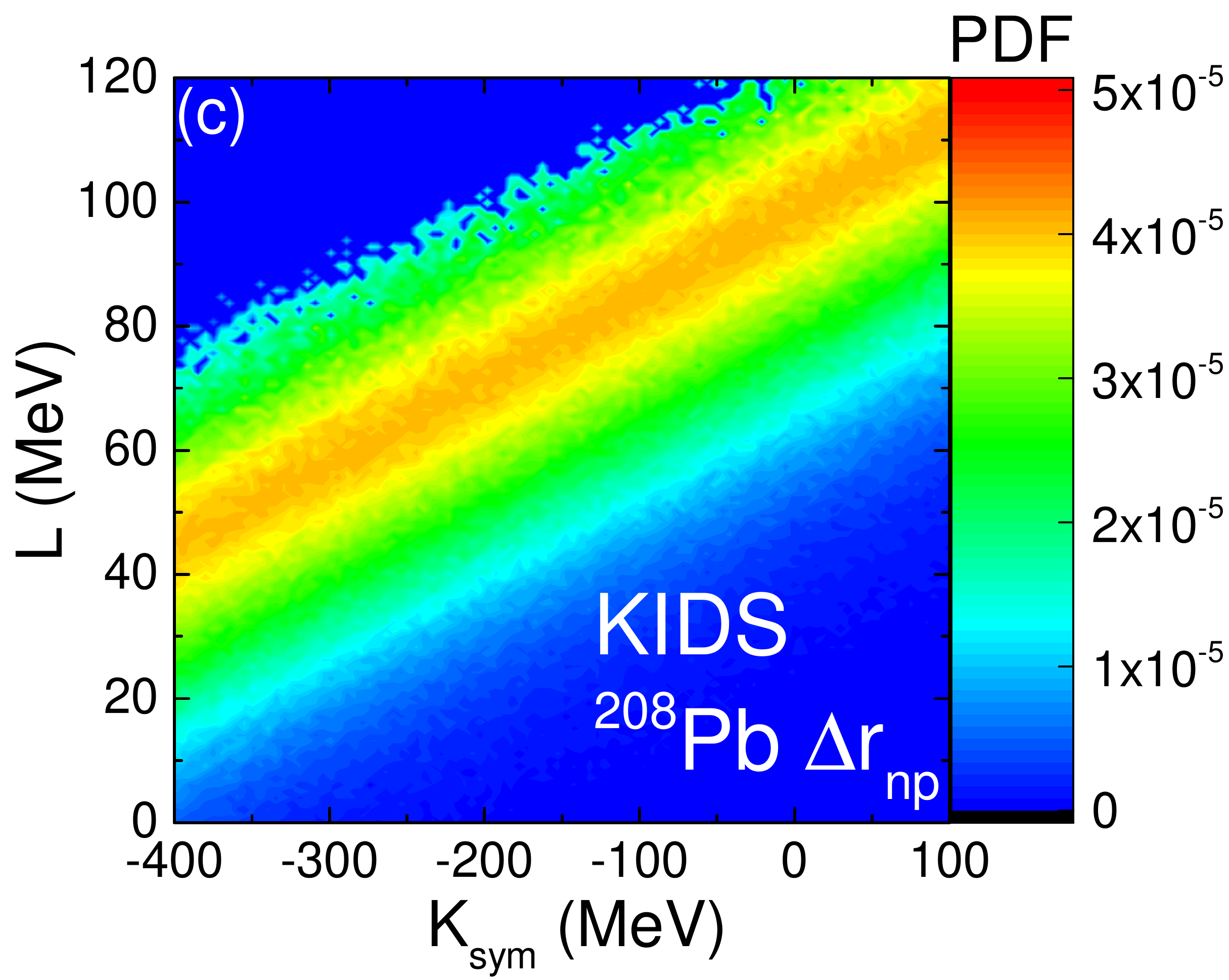}
	\includegraphics[scale=0.16]{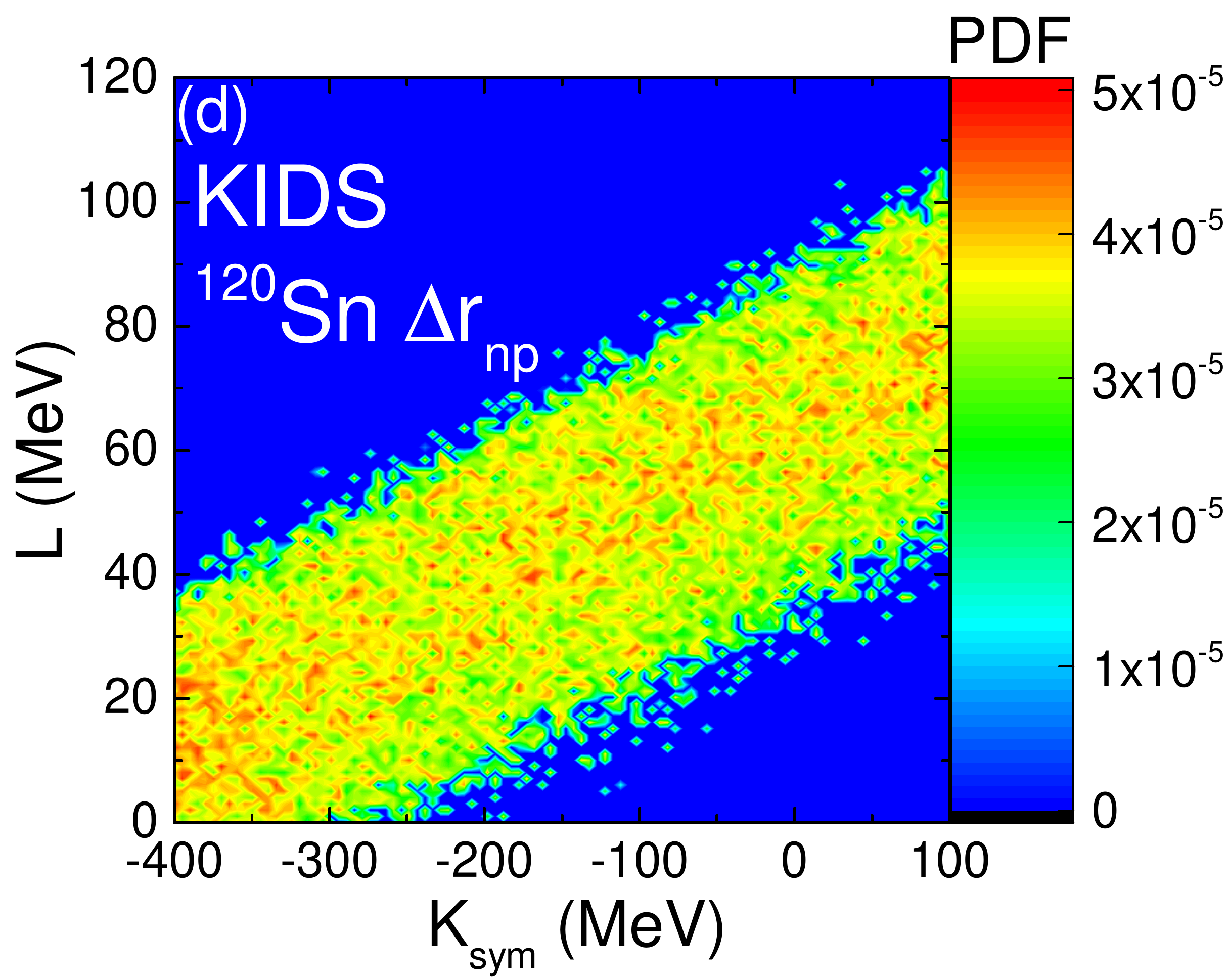}
	\caption{(Color online) Upper: Correlated PDFs between $L$ and $E_{sym}^0$ with $K_{sym}$ varied as an independent variable from the Bayesian inference of the neutron-skin thickness in $^{208}$Pb and $^{120}$Sn based on the KIDS model; Lower: Correlated PDFs between $L$ and $K_{sym}$ from Bayesian inference of the neutron-skin thickness in $^{208}$Pb and $^{120}$Sn based on the KIDS model. Taken from Ref.~\cite{PhysRevC.105.044305}.}\label{L-Ksym}
\end{figure}

Previously, we have observed an anti-correlation between $L$ and $E_{sym}^0$ under the constraint of the neutron-skin thickness based on the standard SHF model, where $K_{sym}$ can be calculated from $L$, $E_{sym}^0$, and other physics quantities based on the SHF energy-density functional. By varying $K_{sym}$ as an independent variable in the KIDS model, we found that the correlated PDFs between $L$ and $E_{sym}^0$ are smeared out under the constraints of $\Delta r_{np}$ for both $^{120}$Sn and $^{208}$Pb, as shown in the upper panels of Fig.~\ref{L-Ksym}. Since additional degrees of freedom are incorporated in this case, the symmetry energy can no longer be parameterized as $E_{sym}(\rho)=E_{sym}^0(\rho/\rho_0)^\gamma$, so the argument in the appendix of Ref.~\cite{PhysRevC.102.044316} is not valid. Interestingly, positive correlations between $L$ and $K_{sym}$ are observed under the constraint of $\Delta r_{np}$, as shown in the lower panels of Fig.~\ref{L-Ksym}.

\begin{figure}[ht]
	\includegraphics[scale=0.16]{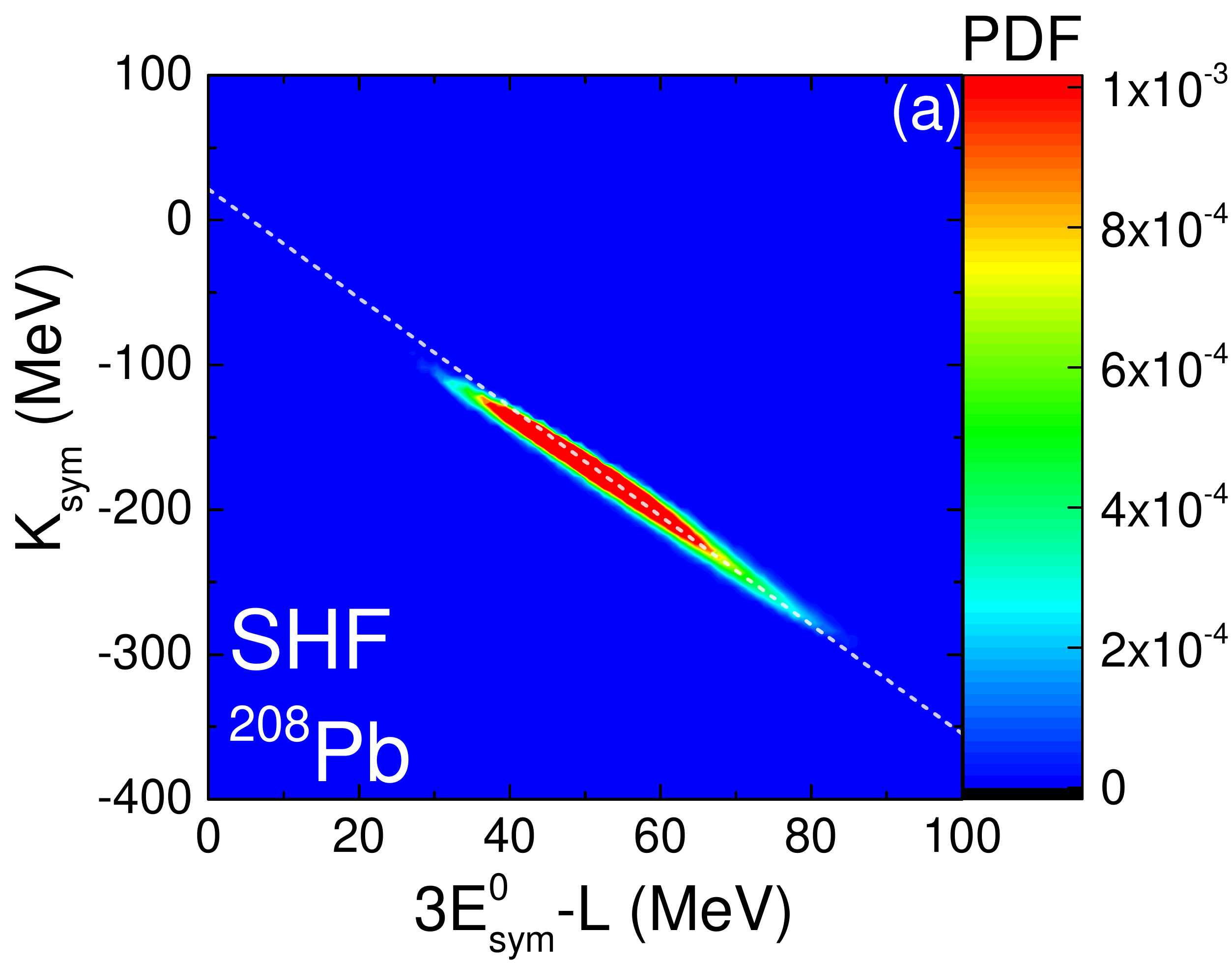}
	\includegraphics[scale=0.16]{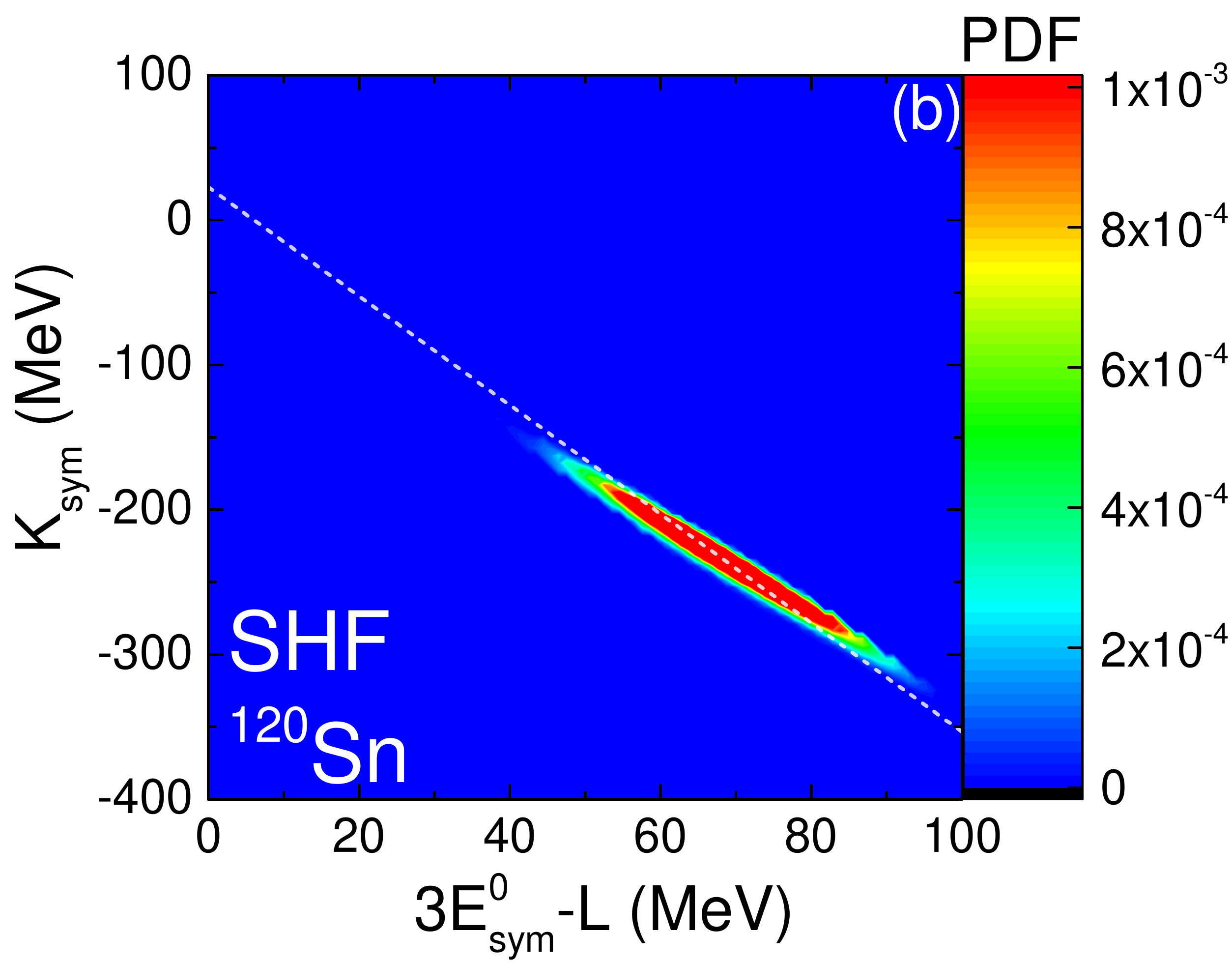}\\
	\includegraphics[scale=0.16]{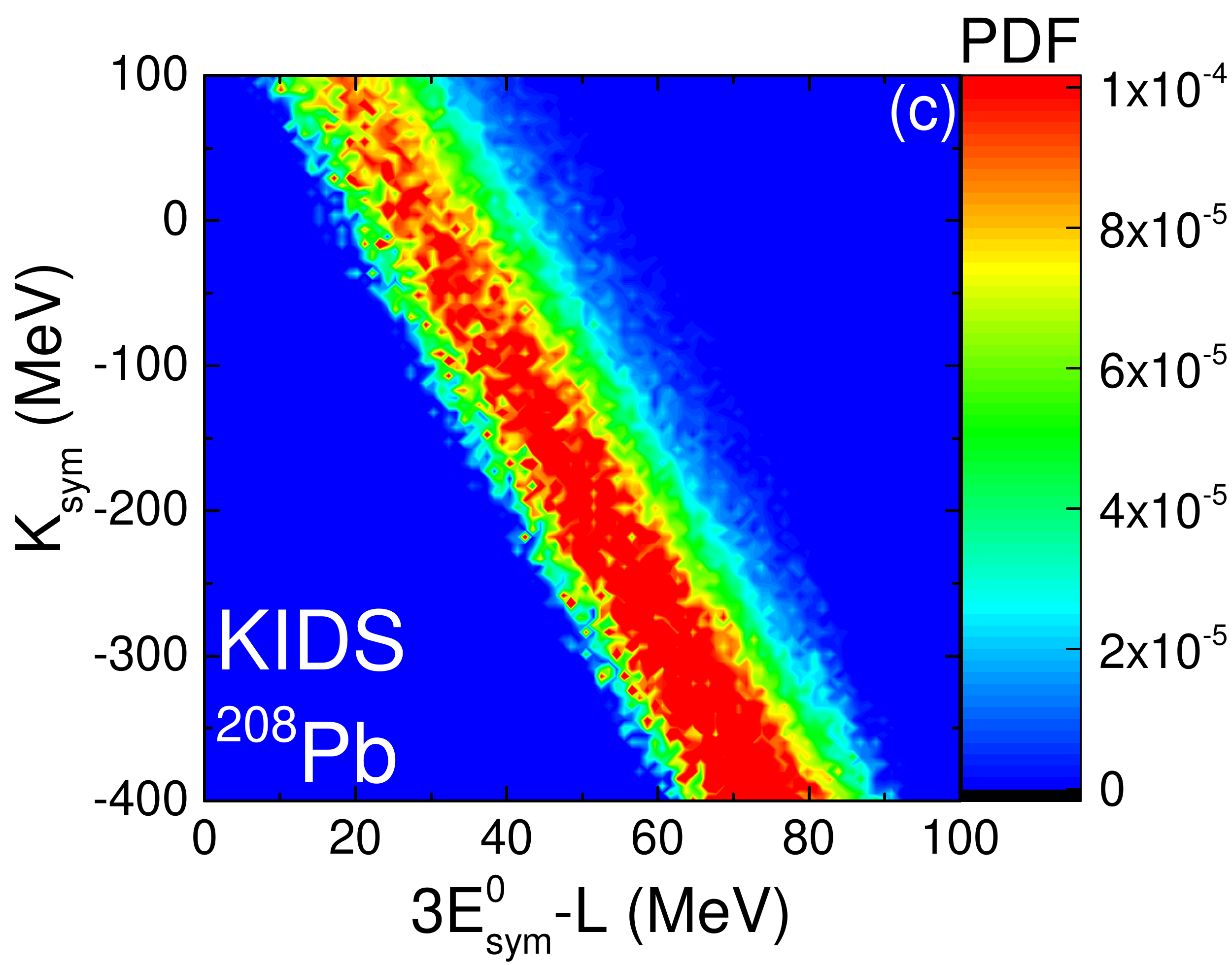}
	\includegraphics[scale=0.16]{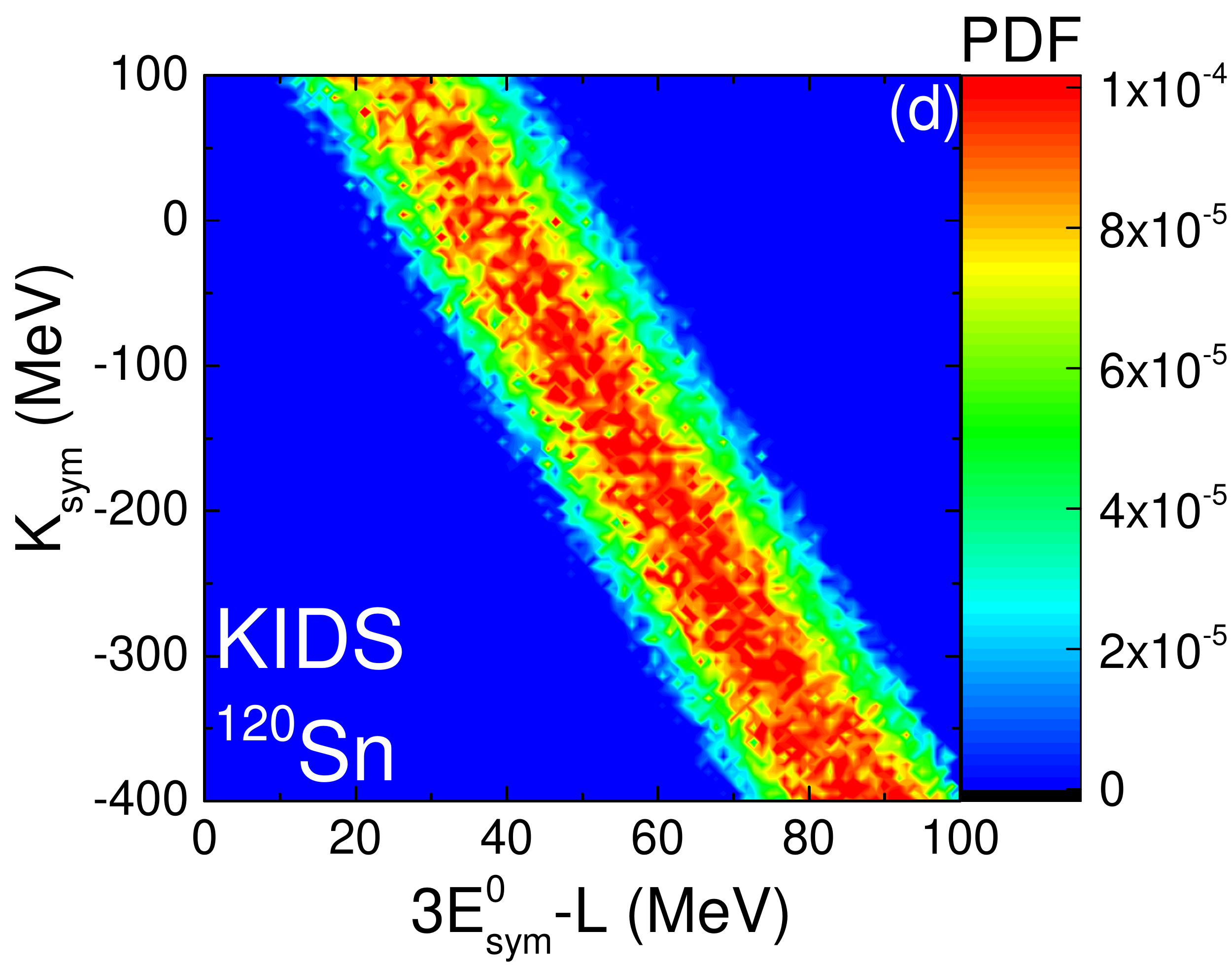}
	\caption{(Color online) Correlated PDFs between $K_{sym}$ and $3E_{sym}^0-L$ from the Bayesian inference of the nuclear structure data of $^{208}$Pb and $^{120}$Sn based on the standard SHF (upper) and the KIDS (lower) model. White dashed lines in the upper panels are intrinsic relations based on the standard SHF model with fixed isoscalar parameters. Taken from Ref.~\cite{PhysRevC.105.044305}.}\label{Ksym-3Esym0-L}
\end{figure}

The linear anti-correlation between $K_{sym}$ and $3E_{sym}^0-L$ has been found to be general in various energy-density functionals~\cite{PhysRevC.96.021302}. In the standard SHF model, the intrinsic anti-correlation relation between $K_{sym}$ and $3E_{sym}^0-L$ at fixed isoscalar parameters is displayed by the white dashed line in the upper panels of Fig.~\ref{Ksym-3Esym0-L}. Under the constraints of the $\Delta r_{np}$ and IVGDR data, both $K_{sym}$ and $3E_{sym}^0-L$ are constrained within a certain range. In the KIDS model, since $K_{sym}$ is an independent variable, there is no such intrinsic relation before being confronted with the experimental data. However, it is seen in the lower panels of Fig.~\ref{Ksym-3Esym0-L} that anti-correlations between $K_{sym}$ and $3E_{sym}^0-L$ are still observed under the constraints of the $\Delta r_{np}$ and IVGDR data, but with a different slope compared with that from the standard SHF model. This shows that the intrinsic anti-correlation between $K_{sym}$ and $3E_{sym}^0-L$ built in the standard SHF model is in some sense reasonable.

\begin{figure}[ht]
	\includegraphics[scale=0.16]{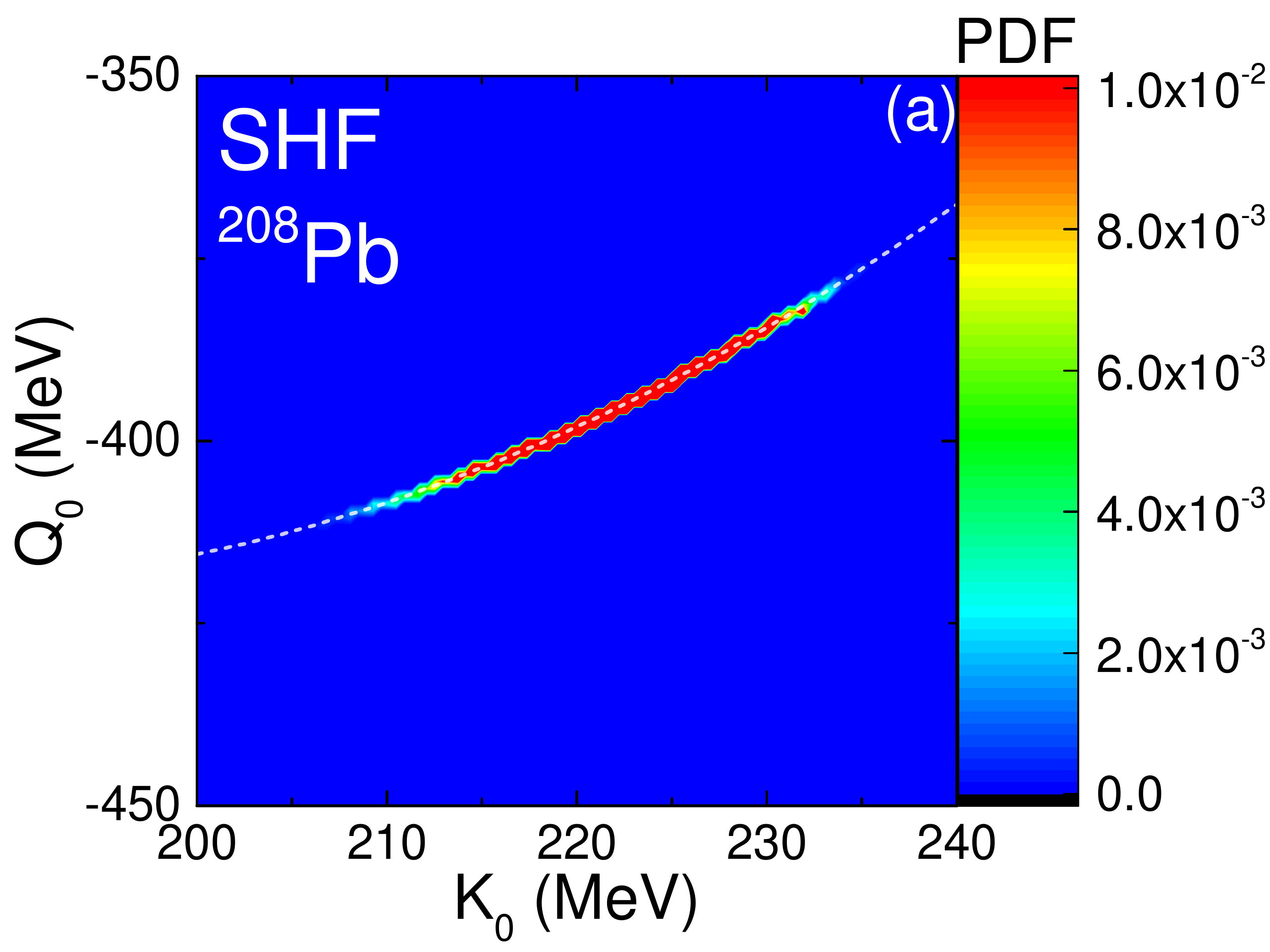}
	\includegraphics[scale=0.16]{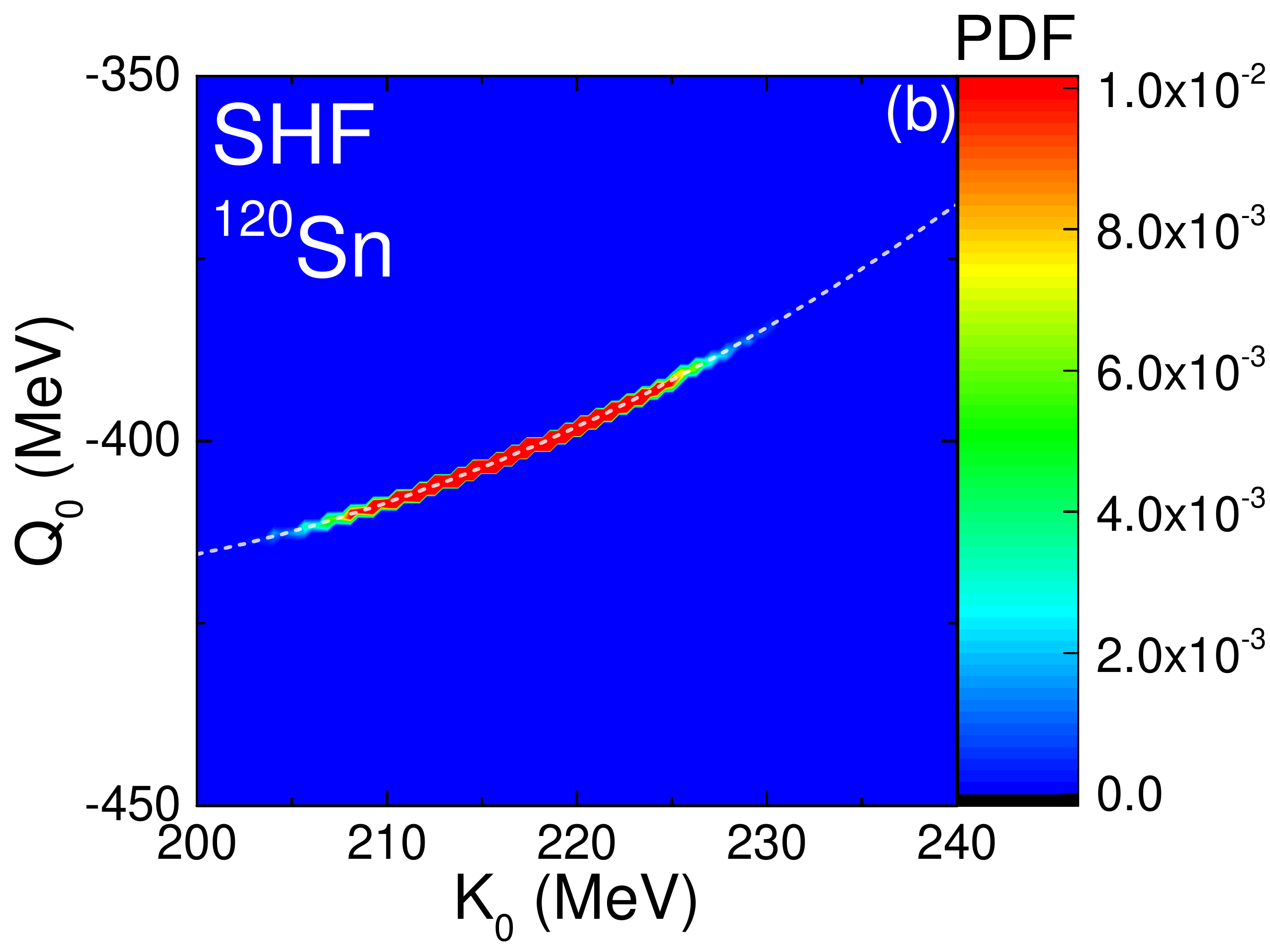}\\
	\includegraphics[scale=0.16]{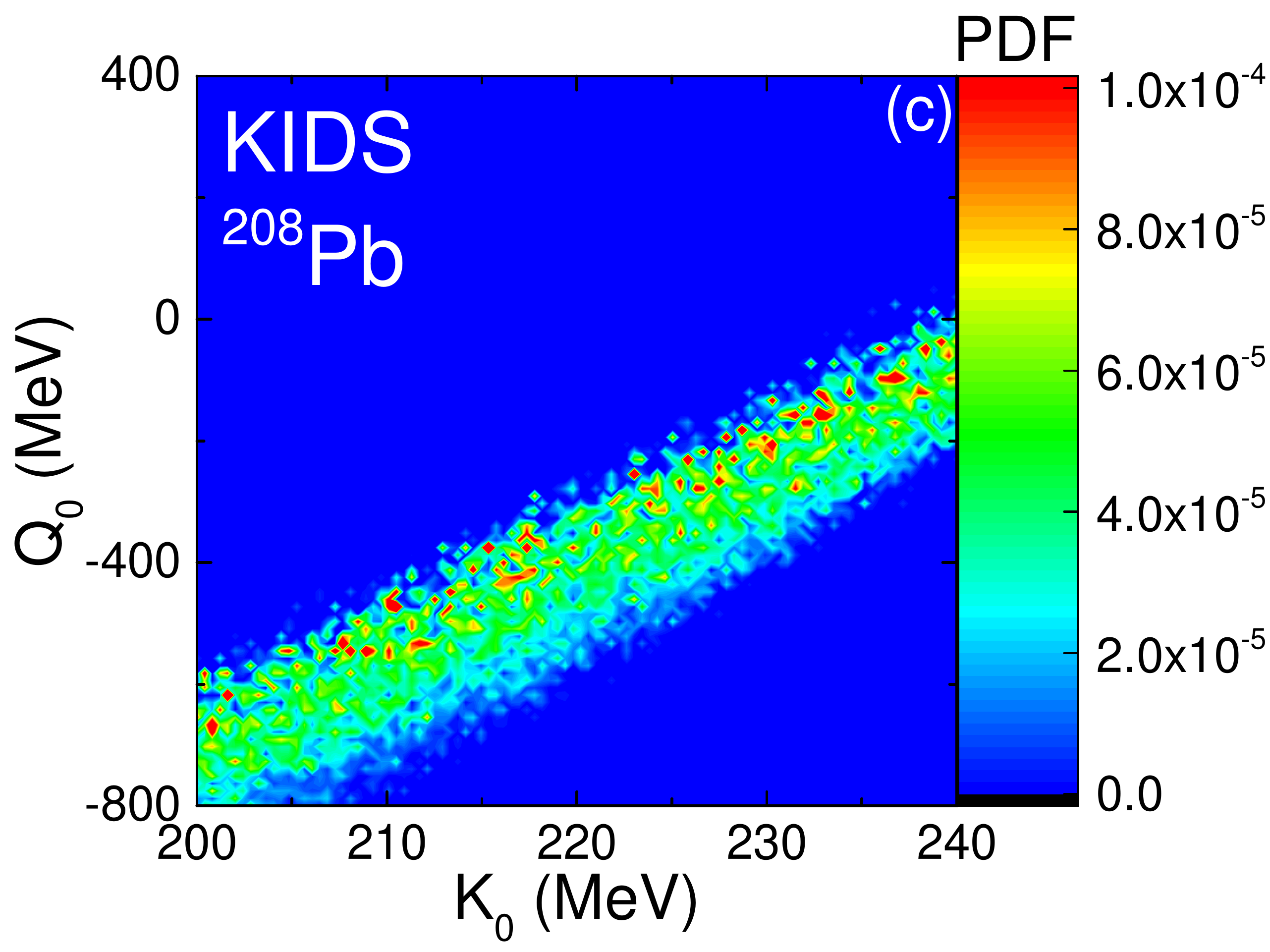}
	\includegraphics[scale=0.16]{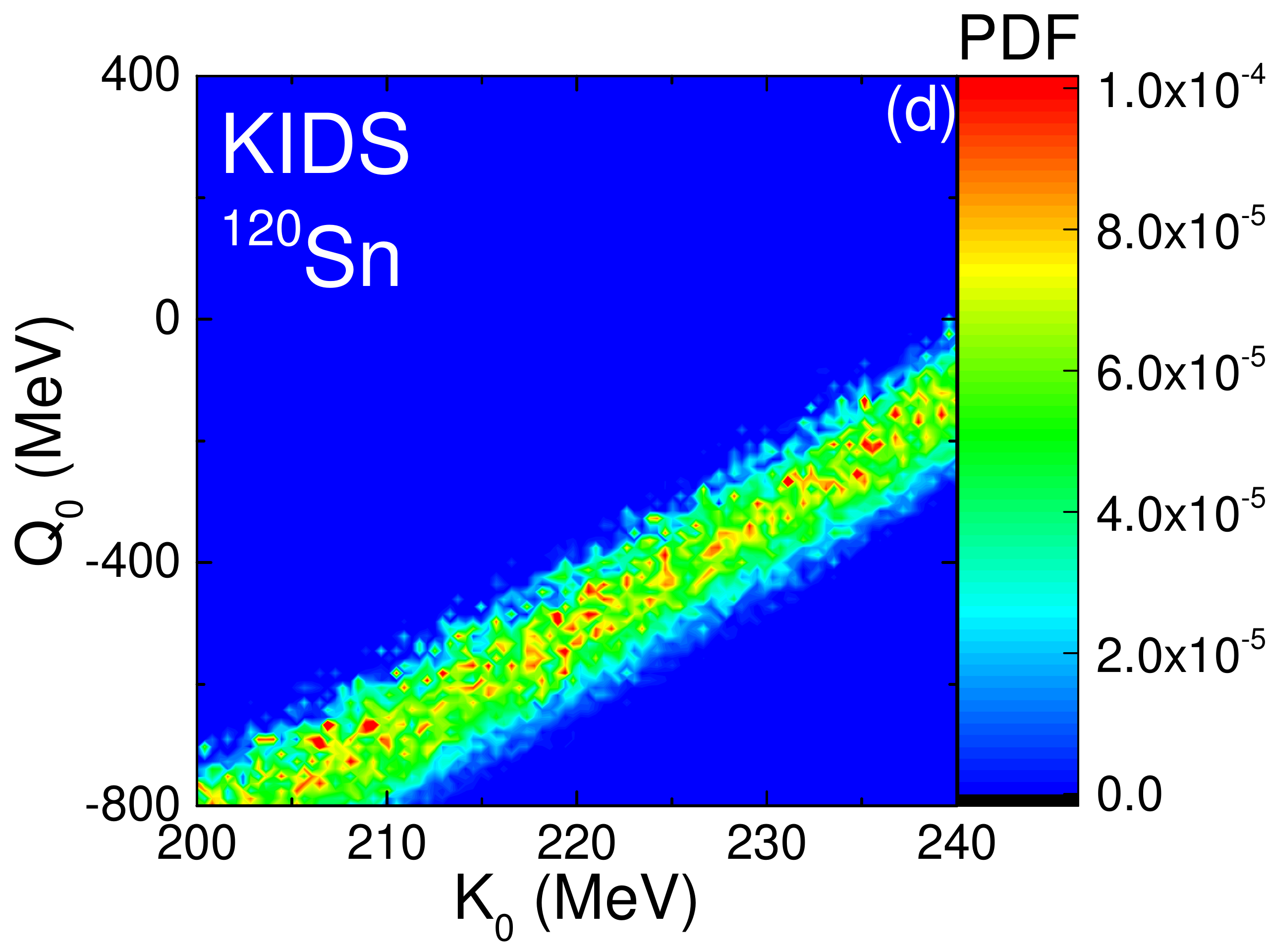}
	\caption{(Color online) Correlated PDFs between $Q_0$ and $K_0$ from the Bayesian inference of the nuclear structure data of $^{208}$Pb and $^{120}$Sn based on the standard SHF (upper) and the KIDS (lower) model. White dashed lines in the upper panels are intrinsic relations based on the standard SHF model with other isoscalar parameters fixed. Taken from Ref.~\cite{PhysRevC.105.044305}.}\label{Q0-K0}
\end{figure}

Previously, we have found that the ISGMR data can help to constrain $K_0$ based on the standard SHF model. The white dashed line in the upper panels of Fig.~\ref{Q0-K0} shows the intrinsic positive correlation relation between $Q_0$ and $K_0$ with other isoscalar parameters fixed in the standard SHF model. Under the constraint of the ISGMR data, $K_0$ is indeed constrained, while $Q_0$ is also constrained within a certain range. Based on the KIDS model where $Q_0$ can be varied as an independent variable, the positive correlation between $Q_0$ and $K_0$ is still seen under the constraint of the ISGMR data, though there are difference between results from the two models. This also shows that the intrinsic positive correlation between $Q_0$ and $K_0$ built in the standard SHF model is in some sense reasonable.

We would like to give some remarks on the different results from different nuclei and based on different models. Since $^{208}$Pb has a larger $\Delta r_{np}$ and an ``effectively'' higher excitation energy of the ISGMR compared with $^{120}$Sn, this leads to a shifted correlated PDF to larger $L$ in Fig.~\ref{L-Ksym} or to smaller $3E_{sym}^0-L$ in Fig.~\ref{Ksym-3Esym0-L}, and to larger $K_0$ in Fig.~\ref{Q0-K0}, for $^{208}$Pb compared to that for $^{120}$Sn. On the other hand, after integrating $K_{sym}$ in the correlated PDFs in the lower panels of Fig.~\ref{L-Ksym} and $Q_0$ in the correlated PDFs in the lower panels of Fig.~\ref{Q0-K0}, one expects that the one-dimensional PDFs of $L$ and $K_0$ are much broader in the KIDS model, compared to the those in Figs.~\ref{L} and \ref{K0} in the standard SHF model. This shows that incorporating higher-order EOS parameters as in the KIDS model may generally weaken the constraints on the lower-order EOS parameters.

\section{Summary and outlook}
\label{sec:summary}

Using the Bayesian analysis, we have presented some highlight results on the constraints of EOS parameters from the data of nucleus resonances and neutron-skin thicknesses based on the standard SHF model as well as its extension. We have discussed the anti-correlation and the positive correlation between $L$ and $E_{sym}^0$ under the constraint of, respectively, the neutron-skin thickness and the IVGDR data, and have shown that we can find a compromise for both the ``PREXII puzzle" and the ``soft Tin puzzle''. With higher-order EOS parameters incorporated as independent variables, while the qualitatively conclusions still hold, we found that the PDFs of lower-order EOS parameters will be broadened and the related correlations between EOS parameters can be modified.

We note that the above compromise for the constraint on the $E_{sym}$ is mainly due to the large $1\sigma$ error for the experimental data, especially for the $\Delta r_{np}$ from electron parity-violating scatterings. The puzzle will be more significant if the experimental error is reduced while the mean value remains unchanged. On the other hand, different observables mainly constrain the behavior of the $E_{sym}$ at different density regions (see, e.g., Ref.~\cite{Lynch:2021xkq} for a recent review). For instance, the IVGDR data mainly constrains the value of the $E_{sym}$ at about $\rho=\rho_0/3$, while the neutron-skin data mainly constrains the slope parameter of the $E_{sym}$ at about $\rho=2\rho_0/3$. We note that at $\rho=2\rho_0/3$, the value of the $E_{sym}$ is tightly constrained by the nuclear masses~\cite{Zhang:2013wna,Danielewicz:2013upa}. While the $E_{sym}$ around these densities can be accurately determined, the extension of the $E_{sym}$ to the saturation density may depend on the energy-density functional. To extend the present studies, we may choose, e.g., $L(2\rho_0/3)$ and/or $E_{sym}(2\rho_0/3)$ as independent parameters instead of that at the saturation density, or including more nuclear structure data such as nuclear masses, in the future.

Generally, the Bayesian approach is a good analysis tool suitable for multi-messengers versus multi-variables, and may help to obtain quantitatively the PDFs of EOS parameters as well as their correlations. On the other hand, the results depend on the energy-density functional and the parameter space. It is the model that builds the relation between parameters and observables, and the Bayesian analysis serves as a good tool to reveal that relation in a proper way. The previous studies are based on the non-relativistic SHF model and its extension, and studies on the relativistic mean-field model are called for to further explore the model dependence on the results. In addition, the data of nucleus resonances and neutron-skin thicknesses mostly constrains the nuclear matter EOS around and below the saturation density. It will be interesting to use the astrophysics data to further constrain the EOS from low to high densities based on a similar framework. Such study is in progress.

\begin{acknowledgments}
JX is supported by the National Natural Science Foundation of China under Grant No. 11922514.
\end{acknowledgments}

\bibliography{bayes}
\end{document}